\documentclass[12pt]{iopart}
\usepackage{graphics,amssymb}

\begin{document}
\title[Stability in a system of globally coupled rotors]
{Stability of thermodynamic and dynamical order in a system of globally coupled rotors}

\author{J. Choi\dag, M Y Choi\ddag\P}
\address{\dag Department of Physics, Keimyung University, Daegu 704-701, Korea}
\address{\ddag Department of Physics, Seoul National University, Seoul 151-747, Korea}
\address{\P Korea Institute for Advanced Study, Seoul 130-722, Korea}

\begin{abstract}
A system of globally coupled rotors is studied in a unified
framework of microcanonical and canonical ensembles. We consider
the Fokker-Planck equation governing the time evolution of the
system, and examine various stationary as well as non-stationary
solutions.  The canonical distribution, describing equilibrium, provides
a stationary solution also in the microcanonical ensemble, which
leads to order in a system with ferromagnetic coupling at low
temperatures.  On the other hand, the microcanonical ensemble
admits additional stationary and non-stationary solutions; the
latter allows dynamical order, characterized by multiple degrees
of clustering, for both ferromagnetic and antiferromagnetic
interactions.  We present a detailed stability analysis of these
solutions: In a ferromagnetic system, the canonical distribution
is observed stable down to a certain temperature, which tends to
get lower as the number of Fourier components of the perturbed distribution 
is increased in the analysis. 
The non-stationary solution remains neutrally stable below 
the critical temperature, indicating inequivalence between the two ensembles.  
For antiferromagnetic systems, all solutions are found to be neutrally stable at all
temperatures, suggesting that dynamical ordering is relatively
easy to observe at low temperatures compared with ferromagnetic systems.
\end{abstract}


\maketitle

\section{Introduction\label{sec:intro}}

The system of sinusoidally coupled oscillators serves as a
prototype model describing various oscillatory phenomena in
nature.  When the coupling is short-ranged, i.e., between nearest
neighbors, the oscillator system describes an array of Josephson
junctions, which has been a subject of extensive studies
\cite{myc}. On the other hand, there are also many systems with
long-range couplings in physics and biology. 
Physiological rhythmic processes may be examples of the latter,
which may be modelled as a system of coupled oscillators with the
range of coupling being  varied, where phase synchronization of
the system is an important issue to be understood~\cite{glass}.
Physical examples are diverse, ranging from self-gravitating and
plasma systems, where the long-range nature of the gravitational or
Coulombic interaction
gives rise to difficulty in understanding the systems. 
A system of globally coupled rotors has thus been proposed and
studied to simulate those systems~\cite{ar}. Here the interaction
range is infinite, with the strength scaled with the system size,
making the system of the mean-field character and amenable to
analytical treatment. In spite of the mean-field nature, however,
the system has turned out to exhibit rich features in dynamical
and statistical properties.

In the canonical ensemble one can find an analytic solution and
the system with the ferromagnetic interaction undergoes an
equilibrium phase transition at a finite critical temperature,
whereas there is no phase transition for the antiferromagnetic
interaction.  On the other hand, direct simulations in the
microcanonical ensemble reveal some interesting features with
remarkable differences with the nature of the interaction.
Specifically, for the ferromagnetic interaction, the system
displays extremely slow relaxation towards the thermodynamic
equilibrium. This slow relaxation, dubbed quasi-stationarity, does
not coincide with predictions in the canonical ensemble, and thus
suggestion has been made that there may exist inequivalence
between canonical and microcanonical ensembles. Such
quasi-stationarity is observed to survive well below the
equilibrium critical temperature and hence has attracted much
attention \cite{ar,lrt,ano,has,mta,plr,yam,zan,ybb}, together with
some controversy \cite{mta}.  In the regime showing
quasi-stationarity it has also been reported that the system
exhibits aging effects and glassy behavior \cite{mta,plr}.  For
the antiferromagnetic interaction the system exhibits a different
type of coherent motion at low temperatures, again only in the
microcanonical ensemble \cite{afrot}: The rotors move in two
groups, called the bi-cluster, for a long time, which is explained
in terms of the statistical equilibrium of the effective
Hamiltonian obtained after averaging out fast variables.

In a recent work we have employed a novel approach that treats the
system in a unified framework of microcanonical and canonical
ensembles \cite{cc}. Starting from the set of Langevin equations
describing dissipative dynamics of a system (canonical ensemble)
and the corresponding Fokker-Planck equation (FPE), we have
pointed out that the nondissipative Hamiltonian dynamics
(microcanonical ensemble) may be described as a limiting case of
the vanishing damping coefficient.  Thereupon we have been able to find a
class of solution for the incoherent phase depending on the
ensemble, some of which are neutrally stable even below the
equilibrium critical temperature.  This neutral stability has then
been suggested to be a plausible physical explanation as to the
origin of the quasi-stationarity observed in numerical
experiments.  In this paper we further extend the stability
analysis of the previous work to the ferromagnetic coherent phase
(with thermodynamic order) and to the system with the
antiferromagnetic interaction. For the latter, we attempt to
provide an alternative view of the bi-cluster phase observed in
the antiferromagnetic system, as dynamical order allowed by the
rotating solution of the FPE. This rotating solution is found to
be neutrally stable down to zero temperature. Furthermore, the
rotating solution can give rise to any degree of clustering, if
the initial condition is appropriately chosen, in addition to
bi-clustering. It would thus be of interest to probe such multi-cluster
motions as tri-clustering, etc., by means of numerical simulations.

This paper is organized as follows: In Sec. {\ref{sec:model}} we
describe how the system of globally coupled rotors can be treated
in a unified framework from the set of Langevin equations and the
corresponding FPE. Various solutions of the FPE are given in Sec.
{\ref{sec:solfp}}. It is shown that multi-cluster solutions
emerge, manifesting dynamical order for the non-stationary
rotating solution of the FPE. Section {\ref{sec:sta}} is devoted
to the stability analysis of the stationary solutions, with
emphasis on the ferromagnetically coherent phase (with single
cluster motion or thermodynamic order). The stability analysis of
the non-stationary solution is presented in Sec.
{\ref{sec:nonsta}}, with a special focus on the antiferromagnetic
case. Finally, a brief summary is given in Sec. {\ref{sec:con}}.

\section{System of Coupled Rotors \label{sec:model}}

We consider a system of $N$ classical rotors,
each of which is described by its phase angle
and coupled sinusoidally to others.
The dynamics of the coupled rotor system is governed by the set of equations of motion
for the phase $\phi_i$ ($i=1,...,N$) of the $i$th rotor:
\begin{equation}\label{eqm}
    M\ddot{\phi_{i}} +\sum_j J_{ij} \sin (\phi_i -\phi_j)=0,
\end{equation}
where $M$ is the inertia of each rotor and
$J_{ij}$ represents the coupling strength between rotors $i$ and $j$.
With the introduction of the canonical momentum $p_i =
M\dot{\phi_i}$, the above equations are transformed into a set of
canonical equations:
\begin{equation}
    \dot{\phi_i}=\frac{\partial \mathcal{H}_N}{\partial p_i},~~ \;
    \dot{p_i}=-\frac{\partial \mathcal{H}_N}{\partial \phi_i}
\end{equation}
with the $N$-particle Hamiltonian
\begin{equation}\label{hamil}
    \mathcal{H}_N=\sum_{i}\frac{\displaystyle{p^{2}_{i}}}{\displaystyle{2M}}
    -\sum_{i<j} J_{ij}\cos (\phi_i-\phi_j),
\end{equation}
on which the microcanonical description is based.

On the other hand, in the canonical description the system is in
contact with a heat reservoir of temperature $T$ and described, in
a most general way, by the set of Langevin equations:
\begin{equation}\label{eqmo}
    M\ddot{\phi_{i}}+\Gamma \dot{\phi_{i}} +\sum_j J_{ij} \sin (\phi_i
    -\phi_j)=\eta_i ,
\end{equation}
where $\Gamma$ is the damping coefficient and the Gaussian white
noise $\eta_i (t)$ is characterized by the average $\langle \eta_i
(t) \rangle =0$ and the correlation $\langle \eta_i (t) \eta_j
(t')\rangle = 2 \Gamma T \delta_{ij} \delta (t{-}t')$.
To derive the corresponding FPE, we write the equations of motion in the form
\begin{eqnarray}
 \dot{\phi_i}&=&\frac{\displaystyle{p_i}}{\displaystyle{M}} \nonumber\\
 \dot{p_i}&=&
  -\frac{\displaystyle{\Gamma}}{\displaystyle{M}} p_i -\sum_j J_{ij}
\sin (\phi_i
    -\phi_j)+ \eta_i. \label{langevin}
\end{eqnarray}
It is then straightforward to derive,  
via the standard procedure~\cite{haken},
the FPE for the probability distribution $P({\phi_i},{p_i},t)$:
\begin{eqnarray}
\!\! \frac{\displaystyle{\partial P}}{\displaystyle{\partial t}}
  &=& - \sum_i \frac{\displaystyle{p_i}}{\displaystyle{M}}\frac{\displaystyle{\partial P}}
 {\displaystyle{\partial \phi_i}}
  + \sum_i\frac{\displaystyle{\partial }}{\displaystyle{\partial p_i}} \nonumber \\
  &\times&\!\left[\frac{\displaystyle{\Gamma}}{\displaystyle{M}} p_i
  + \sum_j J_{ij} \sin (\phi_i-\phi_j)
  + \Gamma T \frac{\displaystyle{\partial }}{\displaystyle{\partial p_i}}\right] P .
\label{fp}
\end{eqnarray}
One may also derive the FPE for the Hamiltonian dynamics, which just reads
Eq. (\ref{fp}) with $\Gamma = 0$. 
While reflecting that Eq. (\ref{eqmo}) with $\Gamma$ set equal to zero reduces to Eq. (\ref{eqm}),
this suggests that Eq. (\ref{fp}) should provide the starting point for both
descriptions: the microcanonical one ($\Gamma =0$) and the
canonical one ($\Gamma \neq 0$).  In particular, the stationary
solution of Eq. (\ref{fp}) is given by the canonical distribution
$P^{(0)}({\phi_i},{p_i}) \propto e^{-\mathcal{H}_N /T}$, 
describing equilibrium, with the very Hamiltonian in Eq. (\ref{hamil}) 
{\em regardless of} $\Gamma$ being zero or not. 
Note, however, that unlike the canonical ensemble where $T$ represents 
the given temperature, in the microcanonical ensemble $T$ still remains 
as an arbitrary parameter.  
In the latter, one may adjust $T$ to the average kinetic energy, which allows 
the interpretation of $T$ as the temperature. 
This prescription thus establishes correspondence between the two ensembles. 
Note also that in the zero-temperature limit ($T\rightarrow 0$), Eq. (\ref{eqmo})
reduces to the Caldirola-Kanai Hamiltonian dynamics \cite{ck},
which needs external driving to have a nontrivial stationary state.

In order to measure a variety of coherence in the system, we conveniently
introduce the generalized order parameter $\Delta^{(\ell)}$ defined by
\begin{equation}\label{gop}
    \frac{1}{N}\sum_i^N  e^{i \ell \phi_i} \equiv \Delta^{(\ell)} \;  e^{i
    \theta_\ell}.
\end{equation}
Apart from the global phase $\theta_\ell$, non-vanishing values of the
order parameter $\Delta^{(\ell)}$ imply that rotors move as clusters, since
rotors separated with phase angle $2 \pi/\ell$ make contributions
to $\Delta^{(\ell)}$.  It thus can be used as a measure of the distribution
of rotors, particularly, the degree of \emph{clustering}.
For instance, a non-vanishing value for $\ell=1$ corresponds to the emergence of
a mono-cluster (or magnetization), that for $\ell=2$ corresponds to bi-cluster
formation (with separation of $\pi$), and so on.
Note that the $\ell=2$ case may be regarded as the analogue of staggered
magnetization in the short-ranged model.
%

\section{Stationary and Non-Stationary Solutions
\label{sec:solfp}}

In the infinite-range limit ($J_{ij}=J/N$ with $N \rightarrow\infty$), 
we use Eq. (\ref{gop}) for $\ell =1$,
\begin{equation}\label{}
    \Delta^{(1)}=\frac{1}{N}\sum_i e^{i (\phi_i-\theta_1)},
\end{equation}
%
and decouple the set of the equations of motion into a single-particle equation
\begin{equation}\label{seqmo}
    M\ddot{\phi_i} + \Gamma \dot{\phi_i}+ J \Delta^{(1)} \sin (\phi_i-\theta_1)
    =\eta_i,
\end{equation}
satisfied by all rotors.  Henceforth we therefore drop the rotor index $i$ in Eq. (\ref{seqmo}), 
which leads to the standard FPE for the single-rotor probability distribution $P(\phi, p,t)$:
\begin{eqnarray}\label{fpc}
  \frac{\partial P}{\partial t}
     =&- &\frac{p}{M}\frac{\partial P}{\partial \phi}
      +J \Delta^{(1)} \sin (\phi -\theta_1)\frac{\partial P}{\partial
      p} \nonumber \\
      &+ &\Gamma \frac{\partial}{\partial p} \left[ \frac{p}{M}
      + T \frac{\partial }{\partial
      p}\right]P.
\end{eqnarray}
In the absence of damping ($\Gamma =0$), this reduces to the FPE
for the microcanonical ensemble:
\begin{equation}\label{fpmc}
  \frac{\partial P}{\partial t}
     =- \frac{p}{M}\frac{\partial P}{\partial \phi}
      +  J \Delta^{(1)} \sin (\phi-\theta_1) \frac{\partial P}{\partial p},
\end{equation}
which is also referred to as the Vlasov equation in some literature \cite{ar,ybb,afrot}.
In terms of this probability distribution, the generalized order parameter is defined to be
\begin{equation}\label{singorder}
     \Delta^{(\ell)}e^{i \theta_\ell}=\langle e^{i \ell\phi} \rangle = \int dp\; d \phi e^{i
     \ell\phi}P(\phi,p,t).
\end{equation}

\subsection{Stationary Solutions}

For the sake of completeness, we briefly review the results for
stationary solutions of the FPE~\cite{cc}. As pointed out for the
general case, both Eqs. (\ref{fpc}) and (\ref{fpmc}) support the
\emph{same} stationary ($\partial P/\partial t =0$) solution:
\begin{equation}\label{candis}
     P^{(0)}(\phi,p) = \frac{1}{\mathcal{Z}} e^{-\mathcal{H}/T}
\end{equation}with the
single-particle Hamiltonian
\begin{equation}\label{spham}
  \mathcal{H}=\frac{p^2}{2M}- J\Delta^{(1)}\cos ({\phi}-\theta_1) ,
\end{equation}
where the overall phase $\theta_1$ manifests the global $U(1)$ symmetry.
It is thus expected that both ensembles exhibit the same equilibrium behavior.

One, however, should recall again that here $T$ is given in Eq.
(\ref{fpc}) (for the canonical ensemble) but remains arbitrary in
Eq. (\ref{fpmc}) (for the microcanonical ensemble). In the
microcanonical ensemble the temperature should be defined as a
measure of the average kinetic energy according to $\langle p^2
\rangle/2M \equiv T/2$. The partition function is determined by
normalization:
\begin{equation}
  \mathcal{Z} = \int dp \int
  \frac{d\phi}{2\pi}~e^{-\mathcal{H}/T}.
\end{equation}
For later use, we first describe some equilibrium properties of
the globally coupled rotors~\cite{ar} through the use of the single-particle model. 
Defining $x\equiv J\Delta^{(1)} /T$ and making use of the expansion
\begin{equation}\label{exp}
    e^{x \cos (\phi-\theta_1)}=\sum^{\infty}_{n=-\infty} I_{n}(x)e^{in(\phi-\theta_1)}
\end{equation}
with $I_{n}(x)$ being the modified Bessel function of the $n$-th order,
we evaluate the partition function as
\begin{equation}\label{parfun}
     \mathcal{Z} =\sqrt{2\pi MT}I_{0}(x).
\end{equation}

We emphasize again that this approach based on the FPE
provides a unified description of microcanonical and canonical
ensembles and both ensembles generate the same equilibrium
behavior, determined by the same distribution $P^{(0)}(\phi,p)$.
Namely, in both ensembles the generalized order parameter in equilibrium is
given by
\begin{eqnarray}
    \Delta^{(\ell)}e^{i\theta_{\ell}} &=&\langle e^{i \ell\phi}\rangle
        = \int dp\int \frac{d\phi}{2\pi} P^{(0)}(\phi,p) e^{i\ell\phi}. 
\label{order}
\end{eqnarray}
With the expansion in Eq. (\ref{exp}) and integration over $\phi$, the
order parameter reads \cite{comph}
\begin{equation}\label{higher}
    \Delta^{(\ell)}=\frac{I_\ell (x)}{I_0 (x)}.
\end{equation}
Note here that $\Delta^{(\ell)}$ has an explicit dependence on
the coherence order parameter $\Delta^{(1)}$ through $x\equiv J\Delta^{(1)} /T$.
For $\ell=1$, describing the emergence of coherence (the mono-cluster as thermodynamic order),
Eq. (\ref{order}) becomes an equation to be solved self-consistently:
\begin{equation}\label{sceqx}
    \frac{T}{J}\,x=\frac{I_1 (x)}{I_0 (x)}.
\end{equation}
This self-consistency equation determines whether the system exhibits coherence:
The ordered phase ($\Delta^{(1)} \neq 0$) emerges when $T/J$ is smaller than the slope
of $I_1 (x)/I_0 (x)$ at $x=0$, which is 1/2.
Accordingly, the ferromagnetic system ($J>0$) undergoes a phase transition at the critical
temperature $T_c =J/2$. In the case of antiferromagnetic coupling ($J < 0$),
on the other hand, Eq. (\ref{sceqx}) becomes $-Tx/|J| = I_1 (x)/I_0 (x)$,
leading to the only solution $x=0$.  It is thus concluded that the antiferromagnetic
system has no phase transition at finite temperatures (no mono-cluster).
It is obvious in Eq. (\ref{higher}) that $\Delta^{(\ell)}$ for higher values of $\ell$
can assume nonzero values only for $\Delta^{(1)}\neq 0$; this implies that only
the ferromagnetic system can develop all degrees of clustering below $T_c$.
This is not surprising since the mono-cluster phase has a $2\pi$
symmetry and therefore invariant under any rotations of multiples
of $2\pi$, which in turn gives rise to nonzero $\Delta^{(\ell)}$.

%
For $\Delta^{(1)}=0$, describing the incoherent phase, the
single-particle Hamiltonian (\ref{spham}) has only the kinetic
energy term, thus reducing the canonical distribution
$P^{(0)}(\phi, p)$ to the Maxwell distribution for both ensembles.
Unlike Eq. (\ref{fpc}), however, Eq. (\ref{fpmc}), the FPE in the
microcanonical ensemble, allows an extra solution of the form
$P^{(0)}(\phi, p)=f_0 (p)$, an arbitrary function of $p$ without
$\phi$-dependence, including the Maxwell distribution as a special
case~\cite{cc}. The only constraint is the normalization, and the
distribution $P^{(0)}(\phi, p)$ uniform in $\phi$ guarantees
$\Delta^{(1)}=0$.
As a result, $\Delta^{(\ell)}$ vanishes for all values of $\ell$ as well,
and no multi-clustering is allowed by this type of stationary solution
present only in the microcanonical ensemble.

\subsection{Rotating Solutions}
In addition to the stationary solutions presented above, the FPE
in the microcanonical ensemble also carries non-stationary solutions
which have some significance for the antiferromagnetic system.
For $\Delta^{(1)} =0$, Eq. (\ref{fpmc}) becomes
\begin{equation}\label{tdfp}
     \frac{\partial P}{\partial t}
     =- \frac{p}{M}\frac{\partial P}{\partial \phi},
\end{equation}
which has a solution of the general form
$P^{(0)}(\phi,p,t)=u(\phi-p\,t/M,p)$. This is a rotating solution
in the sense that the phase grows continuously with time with a
continuous frequency spectrum ($\omega \propto p/M$). Requiring
periodicity in $\phi$, we write
\begin{equation}\label{non-sta}
    P^{(0)}(\phi,p,t)=\sum_k e^{ik(\phi -\frac{p}{M}t)}F_k (p),
\end{equation}
where $F_k (p)$ is an arbitrary function of $p$ satisfying
$F_{\pm 1}(p)=0$ due to the condition $\Delta^{(1)} =0$.
The generalized order parameter for this solutions is computed according to
\begin{eqnarray}\label{himom}
    \Delta^{(\ell)} e^{i\theta_\ell}&=&\int dp\int
   \frac{\displaystyle{d\phi}}{\displaystyle{2\pi}}\sum_k e^{i \ell \phi}
   e^{ik(\phi -\frac{p}{M}t)}F_k (p)\nonumber  \\
   &=&\int dp \; e^{+i\ell\frac{p}{M}t}F_{-\ell} (p),
\end{eqnarray}
which shows that the higher-order moment $\Delta^{(\ell)}$ in
general does not vanish unless $F_{-\ell}(p)=0$. Thus far there is
no difference between the ferromagnetic and the antiferromagnetic
couplings. As will be shown later, however, the stability of the
rotating solution differs substantially, depending on the nature
of the interaction. The rotating solution exists only for
$\Delta^{(1)} =0$, regardless of whether the system is in
equilibrium or not. While such an incoherent phase appears only at
high temperatures in the ferromagnetic case, $\Delta^{(1)}$
remains always zero at all temperature ranges in the
antiferromagnetic case. Moreover, the rotation frequency gets
higher as the order of the moment increases. This suggests that at
low temperatures where thermal fluctuations are small, the phases
with non-vanishing higher moments (high degrees of clustering) are
easier to observe in the antiferromagnetic system than in the
ferromagnetic one. In fact, this is precisely what has been seen
in recent numerical simulations, which reported the bi-cluster
phase in the antiferromagnetic system at very low
temperatures~\cite{afrot}.  The bi-cluster state, with two
clusters separated by angle $\pi$, may be obtained with suitable
choices of $F_k(p)$ in Eq. (\ref{non-sta}). For example, with the
choice $F_{2k}(p)=F(p)$ and $F_{2k+1}(p)=0$, we
obtain~\cite{period}
\begin{equation}\label{pi1}
     P^{(0)}(\phi,p,t)=\pi
     F(p)\left[\delta\left(\phi{-}\frac{p}{M}t\right)+\delta\left(\phi{-}\frac{p}{M}t+\pi\right)\right].
\end{equation}
For another choice, say $F_{2k}(p)= (-1)^k F(p)$ and $F_{2k+1}(p)=0$, one obtains
\begin{equation}\label{pi2}
     P^{(0)}(\phi,p,t)=\pi F(p)\left[\delta\left(\phi{-}\frac{p}{M}t{+}\frac{\pi}{2}\right)
                     +\delta\left(\phi{-}\frac{p}{M}t{-}\frac{\pi}{2}\right)\right].
\end{equation}
All these are shown to be neutrally stable in Sec. \ref{sec:nonsta}.

Equation (\ref{himom}) further indicates that there can exist higher-order
multi-cluster phases (for $\ell = 2, 3, ...$) as well,
if appropriate choices for $F_k (p)$ are made.
Recall again that the multi-cluster phase does not occur for stationary solutions since
$\Delta^{(\ell)} =0$ for time-independent solutions.
In other words, the multi-cluster must rotate with the frequency
higher as the number of clusters grows;
this suggests that the multi-cluster with large $\ell$ should be difficult to observe.

\section{Stability of Stationary States}\label{sec:sta}

In the previous work \cite{cc}, we have already shown that the stability of the
incoherent phase depends on the solutions of the FPE, providing a plausible explanation
as to the physical origin of the quasi-stationarity. We now extend the
analysis further to include the case of the coherent phase.  For this purpose, we
write the FPE, setting $\Delta^{(1)}\equiv \Delta$ and $\theta_1 \equiv \theta$, in the form
\begin{equation}\label{fps}
  \frac{{\partial P}}{{\partial t}}
    =- \frac{{p}}{{M}}\frac{{\partial P}}{{\partial \phi}}
     + J \Delta \sin (\phi -\theta)\frac{{\partial P}}{{\partial p}}
     + \Gamma \frac{{\partial }}{{\partial p}}\left[ \frac{{p}}{{M}}
     + T \frac{{\partial }}{{\partial p}}\right] P,
\end{equation}
To probe the stability,
we add a small perturbation to write
\begin{equation}\label{ex1}
    P(\phi,p,t)=P_0 (\phi,p,t)+f(\phi,p,t)
\end{equation} and accordingly
\begin{eqnarray}\label{ex2}
    \Delta (t)&=&\Delta_0 (t)+\Delta_1 (t) \nonumber \\
              &=&\int dp \int \frac{d\phi}{2\pi} e^{i (\phi-\theta)}
              \left[ P_0 (\phi,p,t)+f(\phi,p,t)
              \right].
\end{eqnarray}
Substituting these into (\ref{fps}), one obtains, to the lowest order,
%
\begin{eqnarray}
\label{stab}
    \frac{{\partial f}}{{\partial t}}
      =&-& \frac{p}{M}\frac{{\partial f}}{{\partial \phi}}
       + J \Delta_1 \sin (\phi -\theta) \frac{{\partial P_0}}{{\partial p}}  \nonumber \\
        &+& J \Delta_0 \sin (\phi -\theta) \frac{{\partial f}}{{\partial p}}
        + \Gamma \frac{{\partial }}{{\partial p}}\left( \frac{{p}}{{M}}
       + T \frac{{\partial }}{{\partial p}}\right) f.
\end{eqnarray}
Since $f(\phi,p,t)$ and $\Delta_1 (t)$ are periodic in $\phi$, one
can Fourier expand them in plane waves:
\begin{equation}\label{fouf}
     f(\phi,p,t)= \sum_k \int d\omega e^{i(k\phi- \omega t)}\tilde{f}_k(p,\omega)
\end{equation}
and
\begin{eqnarray}\label{foudel}
      \Delta_1 (t)&=&\int dp {\frac{d\phi}{2\pi}}e^{i(\phi-\theta)} f(\phi,p,t) \nonumber \\
      &=&\int d\omega e^{-i\omega t}\int dp \tilde{f}_{-1}(p,\omega),
\end{eqnarray}
where the integration over $\phi$ has been performed. Note here that the
perturbed order parameter is proportional only to $\tilde{f}_{-1}(p,\omega)$
(or to $\tilde{f}_{+1}(p,\omega)$ if the order parameter has been defined to be
$ \Delta =\langle e^{-i(\phi-\theta)}\rangle$).
Inserting these expressions into Eq. (\ref{stab}) and collecting coefficients of
$e^{i(k\phi -\omega t)}$, one finds the relations satisfied by the Fourier coefficients
$\tilde{f}_k(p,\omega)$.

In the case of ferromagnetic coupling, the coherent phase ($\Delta_0 \neq 0$)
arises at temperatures below $T_c$, regardless of the presence of damping.
The stationary solution in Eq. (\ref{candis}) can be written,
with the help of Eqs. (\ref{exp}), (\ref{parfun}), and (\ref{higher}), in the form
\begin{eqnarray}\label{cosol}
    P_0 (\phi, p)
    &=& f_M (p)\sum^{\infty}_{n=-\infty} \frac{I_{n}(x)}{I_0 (x)}\, e^{in(\phi-\theta)} \nonumber \\
    &=& f_M (p)\sum^{\infty}_{n=-\infty} \Delta^{(n)}(x) \,e^{in(\phi-\theta)},
\end{eqnarray}
where $f_{M}(p) \equiv (2 \pi MT)^{-1/2} \exp (-p^2/2MT)$ is the Maxwell distribution
and $x \equiv J\Delta_0 /T$ as before.
When $x=0$, the above equation simply reduces to the Maxwell distribution,
which is stable at temperatures above $T_c$.
Our concern now is how the coherent phase gets its stability as
the temperature is lowered below the critical temperature.
Putting Eqs. (\ref{foudel}) and (\ref{cosol}) into Eq. (\ref{stab}), one
obtains the following equation for the Fourier coefficients $\tilde{f}_k (p,\omega)$:
\begin{eqnarray}
   \left(\omega - \frac{kp}{M}\right)\tilde{f_k}
     &-&\frac{J\Delta_0}{2} \frac{\partial }{\partial p}({\tilde f}_{k-1}
    -{\tilde f}_{k+1})
    -i  \Gamma \frac{{\partial }}{{\partial p}}\left( \frac{{p}}{{M}}
       + T \frac{{\partial }}{{\partial p}}\right) \tilde{f}_k \nonumber\\
       &=&\frac{J}{2} [\Delta^{(k-1)}(x)- \Delta^{(k+1)}(x)] f'_M (p)
         \int\!dp'\tilde{f}_{-1} .
\label{fuleq}
\end{eqnarray}

We note here that the emergence of coherence contributes to the
off-diagonal term in Eq. (\ref{fuleq}) and to the appearance of
higher-order generalized order parameters, making the stability analysis non-trivial.
When $\omega - {kp}/{M} = 0$, we have a continuous spectrum,
and for $\Gamma =0$, Eq. (\ref{fuleq}) becomes
\begin{equation}
    \frac{J\Delta_0}{2} \frac{\partial }{\partial p}{\tilde f}_{k-1}
    =\frac{J}{2} \Delta^{(k-1)}(x)f'_M (p)
        \left[\int\!dp'\tilde{f}_{-1}\right].
\end{equation}
It is easy to show by direct substitution that this equation has a solution of the form
\begin{equation}
    \tilde{f}_k (p,\omega)=\left\{%
\begin{array}{ll}
    f_M (p)h_k (\omega), & \hbox{for $k\neq \pm 1,0, -2$} \\
    0, & \hbox{otherwise.} \\
\end{array}%
\right.
\end{equation}
It is of interest to note this is also the solution for $\Gamma \neq 0$ as well,
since the term including $\Gamma$ vanishes for the Maxwell distribution.
For $\omega - {kp}/{M} \neq 0$, we have a discrete spectrum and may not solve the equation
for the general case.
Still we may proceed further if we take the phase-only perturbation,
namely, $f(p,\phi,t) =f_M (p) h(\phi,t)$.
We then have the Fourier coefficient $\tilde{f}_k (p,\omega)= f_M (p) h_k (\omega) $
with $h_k (\omega)$ being the Fourier coefficient of $h(\phi,t)$, which
in turn gives $(\partial /\partial p){\tilde f}_{k}(p,\omega)=f'_M (p) h_k (\omega)$
and $\int dp {\tilde f}_{k}(p,\omega)=h_k (\omega)$.
Further, the term including $\Gamma$ vanishes identically in this case.
Dividing Eq. (\ref{fuleq}) by $\omega -{kp}/{M}$ and integrating over $p$,
we obtain
\begin{eqnarray}\label{hk}
    h_k (\omega)=
    &-& [\Delta^{(k-1)}(x)- \Delta^{(k+1)}(x)]
     \chi_k (\omega)\tilde{h}_{-1}(\omega) \nonumber \\
     &+& 2 \Delta_0 \chi_k (\omega)[h_{k+1}(\omega)-h_{k-1}(\omega)],
\end{eqnarray}
where we have introduced the $k$-dependent response function
\begin{equation}\label{chik}
    \chi_k (\omega)=\frac{J}{2}\int dp \frac{f'_M (p)}{\omega +kp/M}
\end{equation}
and used Eq. (\ref{sign}).
Some properties of this response function, which is frequency-dependent,
are discussed separately in the Appendix.
For $x \ne 0$, the recursion relation for the modified Bessel functions \cite{as}:
\begin{equation}\nonumber
    I_{k-1}(x)- I_{k+1}(x)=\frac{2k}{x}I_k (x)
\end{equation}
leads Eq. (\ref{hk}) to take the form
\begin{equation}
h_k -2 \Delta_0\chi_k (\omega)(h_{k+1}-h_{k-1})=
\frac{2k}{x}\Delta^{(k)}(x)\chi_k (\omega)h_{-1},
\end{equation}
which
needs to be solved.
For $k=0$, from Eqs. (\ref{hk}) and (\ref{c0}), we find $h_0 =0$,
implying the absence of a constant term in the perturbation.
Noting $\Delta^{(k)}(x)=\Delta^{(-k)}(x)$ and
$\chi_k (\omega) =-\chi_{-k}(\omega)$, we write the difference
equation in the matrix form:
\begin{equation}\label{lam}
    \Lambda \left(%
\begin{array}{c}
  h_{-1} \\
  h_{-2} \\
  h_{-3} \\
  ... \\
\end{array}%
\right)=0
\end{equation}
with the matrix
\begin{equation}\label{all}
\Lambda=
\left(%
\begin{array}{cccc}
  1+(2/x)\Delta^{(1)}\chi_1 & -\Delta_0 \chi_1 & 0 & ... \\
  \Delta_0 \chi_2 +(4/x)\Delta^{(2)}\chi_2 & 1 & -\Delta_0 \chi_2& ... \\
  (6/x)\Delta^{(3)}\chi_3 & \Delta_0 \chi_3& 1 & ... \\
  ... & ... & ... & ... \\
\end{array}%
\right).
\end{equation}
Here we have included the terms with only negative $k$ values,
reflecting that all order parameters are defined by Eq.
(\ref{order}). In order to have non-trivial solutions for
$\vec{h}=(h_{-1},h_{-2},h_{-3},...)$, one should have the
vanishing determinant:
\begin{equation}\label{det}
     \varepsilon(\omega)\equiv \textrm{det} \, \Lambda=0.
\end{equation}

Let us first consider the limit $\Delta_0 \rightarrow 0$ or $x=J\Delta_0 /T
\rightarrow 0$, corresponding to the incoherent phase. In this
limit all the off-diagonal terms vanish, since $I_0 (x)\rightarrow 1$
and $I_n (x) \rightarrow (x/2)^n $ so that $\Delta^{(n)}\rightarrow (x/2)^n$.
Equation (\ref{lam}) then becomes
\begin{equation}
    \left(%
     \begin{array}{cccc}
     1+\chi_1 & 0 & 0 & ... \\
     0 & 1 & 0& ... \\
     0 & 0& 1 & ... \\
     ... & ... & ... & ... \\
    \end{array}%
   \right)
   \left(%
    \begin{array}{c}
     h_{-1} \\
     h_{-2} \\
     h_{-3} \\
     ... \\
    \end{array}%
   \right)=0,
\end{equation}
which leads to
\begin{equation}
\label{zeroco}
    1+\chi_1 (\omega)=1+\chi(\omega) \equiv 1+\frac{J M}{2}\tilde{\chi}(\omega)=0
\end{equation}
for non-vanishing $h_{-1}$, while all other $h$'s are zero. 
The detailed analytic properties of the reduced response function
$\tilde{\chi}_(\omega) \equiv (2/JM)\chi_(\omega)$, with the
complex frequency $\omega=\omega_r +i\omega_i$, are presented in
the Appendix. Equation (\ref{zeroco}) describes the condition for
the incoherent phase with the Maxwell distribution, which is
stable/unstable above/below $T_c$ \cite{cc}. For $x \ne 0$, in
principle we have to solve Eq. (\ref{det}) including all the terms
in Eq. (\ref{all}). Since this is very formidable, we instead
consider just a few terms to explore how the stability of the
solution changes. To this end, we write $\varepsilon(\omega) \,
\approx \varepsilon^{(m)}(\omega)$, the determinant obtained when
the first $m$ Fourier components are kept.  With only the first Fourier
component $h_{-1}$ considered, Eq. (\ref{det}) obtains the form
\begin{eqnarray}\label{nonzeroco}
    \varepsilon^{(1)}(\omega)&=&1+\frac{2}{x}\Delta^{(1)}\chi_1
     (\omega)\nonumber \\
     &=&1+\frac{2T}{J}\chi_(\omega) \nonumber \\
     &=&1+ TM \tilde{\chi}(\omega)=0.
\end{eqnarray}
for which Eqs. (\ref{iplus}) to (\ref{izero}) yield $\omega_i =0$
as the only solution. Comparison of Eq. (\ref{nonzeroco}) with Eq.
(\ref{zeroco}) shows the correspondence $T=J/2=T_c$; this
indicates that the solution is neutrally stable at the critical
temperature, below which coherence develops. Including the next
component $h_{-2}$, one has
\begin{equation}
     \varepsilon^{(2)}(\omega)=\varepsilon^{(1)}(\omega)
     + \Delta_0 (\Delta_0 + \frac{4}{x} \Delta^{(2)})\chi_1(\omega)
    \chi_2 (\omega)=0,
\end{equation}
which, with $\Delta^{(1)}=\Delta_0$ and $\chi_k (\omega)= \chi (\omega /k)/k$,
becomes
\begin{equation}\label{2term}
    \varepsilon^{(2)}(\omega)=1+T\tilde{\chi}(\omega)
         +\frac{T^2}{8}\left[x^2 + 4x \frac{I_2 (x)}{I_1(x)}\right]
          \tilde{\chi}(\omega)\tilde{\chi}(\omega/2)=0.
\end{equation}
Since $\tilde{\chi}(\omega)$ and $\tilde{\chi}(\omega/2)$ have the same pole structure,
the real and the imaginary parts of Eq. (\ref{2term}) read
\label{allequations}
\begin{eqnarray}\
   \textrm{Re}\,\varepsilon^{(2)}(\omega) &\equiv&
    1+ T\textrm{Re} \tilde{\chi}(\omega)
      +\frac{T^2}{8} \left[x^2 + 4x \frac{I_2 (x)}{I_1(x)}\right]  \nonumber \\
     & & ~~~~ \times\left[
    \textrm{Re}\tilde{\chi}(\omega)\textrm{Re}\tilde{\chi}(\omega/2)
    -\textrm{Im}\tilde{\chi}(\omega)\textrm{Im}\tilde{\chi}(\omega/2)
    \right]=0 \label{equationa} \\
    \textrm{Im}\, \varepsilon^{(2)}(\omega) &\equiv&
    T \textrm{Im}\tilde{\chi}(\omega)
       +\frac{T^2}{8} \left[x^2 + 4x \frac{I_2 (x)}{I_1(x)}\right]  \nonumber \\
     & & ~~~~ \times\left[
    \textrm{Im}\tilde{\chi}(\omega)\textrm{Re}\tilde{\chi}(\omega/2)
    +\textrm{Re}\tilde{\chi}(\omega)\textrm{Im}\tilde{\chi}(\omega/2)
    \right]=0. \label{equationb}
\end{eqnarray}
In the Appendix it is shown that $\omega_r =0$ is a solution of
$\textrm{Im} \tilde{\chi}(\omega)=0$, implying that this is also a solution of
$\textrm{Im}\, \varepsilon^{(2)}(\omega)=0$.
For $\omega_i >0 $, Eq. (\ref{equationa}) becomes
\begin{equation} \label{46}
    f(y)-1=\frac{1}{8}\left[x^2 + 4x \frac{I_2 (x)}{I_1(x)}\right]f(y)f(y/2)
\end{equation}
with $y \equiv \omega_i \sqrt{M/2T}$.
As $y$ increases from zero to arbitrarily large values,
the left-hand side of Eq. (\ref{46}) decreases monotonically from zero to $-1$
while the right-hand side is positive-definite for $y>0$.
This suggests that there is no solution for $\omega_i >0$ to make the system unstable.
For $\omega_i =0$, which corresponds to the neutral stability,
we have $\textrm{Re} \tilde{\chi}(\omega)=-1/T$ and thus
Eq. (\ref{equationa}) reads
\begin{equation}
    x^2 + 4x \frac{I_2 (x)}{I_1(x)}=0,
\end{equation}
which leads to $x=0$ for the critical case.
For $\omega_i =- |\omega_i|< 0$, for which the system becomes stable,
Eq. (\ref{equationa}) takes the form
\begin{equation}\label{sol2}
    g(y)-1=\frac{1}{8} \left[x^2 + 4x \frac{I_2 (x)}{I_1(x)}\right] g(y)g(y/2).
\end{equation}
As shown in the Appendix, $g(y) $ is a monotonically increasing
function of $y$ from unity to infinity, and accordingly, $g(y)g(y/2)$ is also
a monotonically increasing function of $y$ in the same domain.
Since the left-hand side of the above equation is monotonically increasing from zero to
arbitrarily large values, Eq. (\ref{sol2}) allows a solution
only for some range of $x$ values. We have determined numerically the
range of $x$ values, in which there exits a solution for $y>0$, to find
\begin{equation}
    x \equiv\frac{J\Delta_0}{T} < x^{(2)}_c \approx 1.32.
\end{equation}
This indicates that the coherent solution is stable only at
temperatures above $T_0$, at which $\Delta_0 (T)$ and $x_c T/J$
meet. We see that the stable region does not extend to the zero
temperature, presumably because we have included only the second
component in our analysis (the first component is trivial). This
may be resolved if one include higher components.  Adding the
third
component $h_{-3}$ leads to the following equation 
\begin{eqnarray}
    \varepsilon^{(3)}(\omega)&=&\varepsilon^{(2)}(\omega)+\frac{1}{24}J^2 \Delta^2_0\left[1+T \tilde{\chi}(\omega)
    + 3T \frac{I_3 (x)}{I_1 (x)}\tilde{\chi}(\omega)\right]
    \tilde{\chi}(\omega/2)\tilde{\chi}(\omega/3) \nonumber \\
   &=& 0
\end{eqnarray}
from which one can perform the similar analysis to find
\begin{eqnarray}\label{3termf}
    & & \left[1 +\frac{x^2}{24} f(y/2)f(y/3) \right][f(y)-1] \nonumber \\
      & &~~~= \frac{1}{8}\left[x^2 + 4x \frac{I_2 (x)}{I_1(x)}
            -\frac{I_3(x)}{I_1(x)}f(y/3)\right]f(y)f(y/2)
\end{eqnarray}
for $\omega_i >0$ and
\begin{eqnarray}\label{3termg}
     & & \left[1 + \frac{x^2}{24} g(y/2)g(y/3)\right][g(y)-1] \nonumber \\
      & &~~~= \frac{1}{8}\left[x^2 + 4x \frac{I_2 (x)}{I_1(x)}
                      -\frac{I_3(x)}{I_1(x)}g(y/3)\right]g(y)g(y/2)
\end{eqnarray}
for $\omega_i <0$. Again, Eq. (\ref{3termf}) does not have a
solution for positive $y$ since the left-hand side is less than
zero while the right-hand side is greater than zero. Equation
(\ref{3termg}) is found to have a solution for $x < x^{(3)}_c
\approx 1.51$. Note here that $x_c$ is increased substantially
once the third component is included, implying that $T_0$, above
which the coherent solution is stable, is lowered. We have
performed this analysis, including up to four components, and
confirmed that this trend persists; this suggests the plausible
conjecture that the coherent solution is stable down to zero
temperature if all the Fourier components are included.

We now turn our attention to the stability of the
antiferromagnetic system for which there is no equilibrium order
($\Delta_0 =0$ or $x=0$). The stability equation reads, for
$J=-|J|$,
\begin{equation}\label{anti}
    1-\frac{|J|}{2}\tilde{\chi}_(\omega)=0,
\end{equation}
which, depending on the sign of $\omega_i$ (with $\omega_r =0$), becomes
\begin{equation}
 \left\{\begin{array}{ll}
      1+ (|J|/2T)f(y) = 0&~\mbox{for} ~ \omega_i > 0 \\
      1+ |J|/2T =0&~\mbox{for} ~ \omega_i = 0 \\
      1+ (|J|/2T) g(y) = 0 &~\mbox{for} ~ \omega_i < 0 .
    \end{array}  \right.
\end{equation}
None of these equations has a solution, since $f(y)$ and $g(y)$ is
positive-definite.
This means that the antiferromagnetic system cannot have self-sustained deviation
in the absence of the perturbation with a discrete spectrum.
On the other hand, with the continuous spectrum $\omega=\omega_r = kp/M$,
the system is neutrally stable at all temperatures.

\section{Stability of Non-Stationary States}\label{sec:nonsta}

As mentioned in Section \ref{sec:solfp}, the non-stationary
solution exists only in the microcanonical ensemble ($\Gamma =0$)
with a continuous frequency spectrum, which can develop spontaneously.
Our concern in this section is the stability of this solution,
especially in the case of the antiferromagnetic interaction.
Equation (\ref{stab}) for stability reads, with $\Delta_0 =\Gamma =0$,
\begin{equation}
    \frac{{\partial f}}{{\partial t}}
      =- \frac{p}{M}\frac{{\partial f}}{{\partial \phi}}
       + J \Delta_1 \sin \phi \frac{{\partial P_0}}{{\partial p}},
\end{equation}
where $P_0$ is given by Eq. (\ref{non-sta}).
The last term in the above equation obtains the form
\begin{eqnarray}
    J \Delta_1 \sin \phi \frac{{\partial P_0}}{{\partial p}}
    &=& \frac{J}{2i}\int d\omega e^{-i \omega t}\left[\int dp'
        \tilde{f}_{-1} (p',\omega)\right](e^{i\phi}-e^{-i\phi}) \nonumber \\
       & & ~~~\times \sum_{k}e^{ik\phi}\frac{\partial}{\partial p}
        \left[\exp\left({-i\frac{kp}{M}t}\right) F_k (p)\right] \nonumber \\
    &=& \frac{J}{2i}\sum_{k}\frac{\partial}{\partial p} \int d\omega
        \left[(e^{i(k+1)\phi}-e^{i(k-1)\phi})\right]\exp
        \left[{-i(\omega+\frac{kp}{M})t}\right] \nonumber \\
       & & ~~~\times F_k (p)\int dp' \tilde{f}_{-1} (p',\omega) \nonumber \\
    &=& \frac{J}{2i}\sum_{k}\int d\omega e^{i(k\phi -\omega t)}
        \frac{\partial}{\partial p}
                \left[{\tilde F}_{k-1}(p,\omega)-{\tilde F}_{k+1}(p,\omega)
                \right]
\end{eqnarray}
with
\begin{equation}\label{efk}
        {\tilde F}_{k}(p,\omega) \equiv F_{k} (p)\int dp'
        \tilde{f}_{-1} (p',\omega-\frac{kp}{M}),
\end{equation}
which leads to the equation for the Fourier coefficients:
\begin{equation}\label{rot}
    \left(\omega - \frac{kp}{M}\right)\tilde{f}_k (p,\omega)
    =\frac{J}{2}\frac{\partial}{\partial p}
     \left[{\tilde F}_{k-1}(p,\omega)-{\tilde F}_{k+1}(p,\omega)
         \right].
\end{equation}
Since we are dealing with the perturbation of the non-stationary
state with a continuous spectrum, the frequency of the perturbation should
satisfy $\omega - kp/M\neq 0$; otherwise, there would be no perturbation at all.
This allows us to divide Eq. (\ref{rot}) by $\omega - kp/M$ and to
integrate over $p$. For $k=-1$, we have
\begin{eqnarray}\label{rosol}
    \int dp'\tilde{f}_{-1} (p',\omega)&=&\frac{J}{2}\int dp
    \left(\omega + \frac{p}{M}\right)^{-1} \nonumber \\
     & \times &\frac{\partial}{\partial p}
     \left[{\tilde F}_{-2}(p,\omega)-{\tilde F}_{0}(p,\omega)
         \right],
\end{eqnarray}
while for $k \ne -1$, $\tilde{f}_k (p,\omega)$ is determined by
$\tilde{f}_{-1} (p',\omega \pm kp/M)$ through Eqs. (\ref{efk}) and
(\ref{rot}). It is thus enough to have non-vanishing
$\tilde{f}_{-1} (p',\omega)$. Now suppose that $\omega =\omega_0$
is a solution of Eq. (\ref{rosol}), i.e.,
$\int dp'\tilde{f}_{-1} (p',\omega) \neq 0$ for $\omega =\omega_0$.
If we write $\int dp'\tilde{f}_{-1} (p',\omega) = K \delta (\omega
-\omega_0)$, then $\int dp'\tilde{f}_{-1} (p',\omega+\frac{2p}{M})
= K \delta (\omega +\frac{2p}{M}-\omega_0)$. Integration over
$\omega$ gives
\begin{eqnarray}
  1+\frac{J M}{2}\int dp \frac{F'_0 (p)}{p+M\omega_0}
     &=& \frac{J M}{2}\int dp \frac{F_{-2} (p)}{(p-M\omega_0)^2} \nonumber \\
     &=& \frac{J M}{2}\int dp \frac{F'_{-2} (p)}{p-M\omega_0},
\end{eqnarray}
where the last line is obtained by integration by parts. Hence the
frequency of a self-sustained oscillation and accordingly the
stability is, similarly to the stationary case [Eq.
(\ref{zeroco})], determined by
\begin{equation}
    1+\frac{J M}{2}\int dp \left[\frac{F'_0 (p)}{p+M\omega_0}
    -\frac{F'_{-2} (p)}{p-M\omega_0} \right]=0.
\end{equation}

The stability condition is thus entirely the same as that of the
stationary case except that we now have two momentum distributions:
From Eqs. (\ref{rechi}) and (\ref{imchi}) with
$M\omega_0 =\tilde{\omega}_r + \tilde{\omega}_i$, we have
\begin{equation}\label{unstaro}
\left\{\begin{array}{ll}
     \displaystyle{\frac{2}{JM}}+\int_{-\infty}^{\infty} dp \,\left[
     \frac{\displaystyle{(p+\tilde{\omega}_r)F'_0 (p)}}
   {\displaystyle{(p+ \tilde{\omega}_r})^2+\tilde{\omega}^2_i}
   -\frac{\displaystyle{(p-\tilde{\omega}_r)F'_{-2} (p)}}
   {\displaystyle{(p- \tilde{\omega}_r})^2+\tilde{\omega}^2_i}\right]=0 \\
   \int_{-\infty}^{\infty} dp \,\frac{\displaystyle{F'_0 (p)}}
    {\displaystyle{(p+ \tilde{\omega}_r)^2+\tilde{\omega}^2_i}}
    -\frac{\displaystyle{F'_{-2} (p)}}
    {\displaystyle{(p- \tilde{\omega}_r)^2+\tilde{\omega}^2_i}}=0,
    \end{array} \right.
\end{equation}
for $\omega_i>0$, for which the system is unstable as the
perturbation grows in time. In the opposite case ($\omega_i< 0$),
the perturbation dies out to make the system stable.  The condition
for this is given by
\begin{equation}\label{staro}
\left\{\begin{array}{ll}
    \displaystyle{\frac{2}{JM}} + \int_{-\infty}^{\infty} dp \,
     \left[\frac{\displaystyle{(p+\tilde{\omega}_r)F'_0 (p)}}
    {\displaystyle{(p+ \tilde{\omega}_r)^2+\tilde{\omega}^2_i}}
    - \frac{\displaystyle{(p-\tilde{\omega}_r)F'_{-2} (p)}}
    {\displaystyle{(p- \tilde{\omega}_r)^2+\tilde{\omega}^2_i}}\right]  \\
    ~~~~~~+2\pi \left[\textrm{Im}F'_0 (-\tilde{\omega})
         -\textrm{Im}F'_{-2} (-\tilde{\omega})\right]=0  \\
     \displaystyle{\tilde{\omega}_i}\int_{-\infty}^{\infty} dp \,
     \left[\frac{\displaystyle{F'_0 (p)}}
    {\displaystyle{(p+ \tilde{\omega}_r)^2+\tilde{\omega}^2_i}}
    -\frac{\displaystyle{F'_{-2} (p)}}
    {\displaystyle{(p- \tilde{\omega}_r)^2+\tilde{\omega}^2_i}}\right] \\
    ~~~~~~+2\pi \left[\textrm{Re}F'_0(-\tilde{\omega})
     -\textrm{Re}F'_{-2}(-\tilde{\omega})\right]=0.
   \end{array} \right.
\end{equation}
Finally, in the neutral case ($\omega_i =0$), the condition simply reads
\begin{equation}\label{neuro}
\left\{\begin{array}{ll}
    \displaystyle{\frac{2}{JM}}+\mathcal{P}\int_{-\infty}^{\infty} dp \,
     \left[\frac{\displaystyle{F'_0 (p)}}{\displaystyle{p+\tilde{\omega}_r}}
     -\frac{\displaystyle{F'_{-2} (p)}}{\displaystyle{p-\tilde{\omega}_r}}\right]=0  \\
    F'_0 (-\tilde{\omega}_r)- F'_{-2} (-\tilde{\omega}_r)=0,
   \end{array} \right.
\end{equation}
where $\mathcal{P}$ stands for the principal part. Our next task
is to determine stability for specific distributions of $F_0 (p)$
and $F_{-2}(p)$. Since most dynamical calculations, for both
ferromagnetic and antiferromagnetic system, have used the so-called
water-bag distribution, we also consider the momenta to be distributed
uniformly in the range $[-\alpha,\alpha]$:
\begin{equation}
    F_0 (p)=\pm F_{-2}(p) =\frac{1}{2\alpha}
\end{equation}
Substitution of $F_0 (p)=\pm
F_{-2}(p)=(2\alpha)^{-1}[\delta(p{+}\alpha)-\delta(p{-}\alpha)]$ into
Eqs. (\ref{unstaro})-(\ref{neuro}), depending on the sign of
$\omega_i$, determines the frequency $\omega_0$.

We first consider the case  $F_0 (p)= F_{-2}(p)$.
From the second equations of Eqs. (\ref{unstaro}) (for $\omega_i >0$) and
(\ref{staro}) (for $\omega_i <0$), we find $\omega_r =0$,
while there is no solution to satisfy the first ones.
When $\omega_i =0$, again there is no solution to satisfy the first equation of
Eq. (\ref{neuro}).  This indicates that there is no self-sustained oscillation in
the system.  Note, however, that the system is neutrally stable as it has a continuous
spectrum ($\omega=kp/M$).
Next, when $F_0 (p)= -F_{-2}(p)$, one finds
  \begin{eqnarray}\label{rof}
     \omega_i& =&\pm\sqrt{\displaystyle{{\frac{J}{M}}-\left({\frac{\alpha}{M}}
   \right)^2}} ,~ \omega_r =0~~\mbox{for}~ \alpha < \alpha_R \\
    \omega_r &=&\pm
    \sqrt{\displaystyle{-{\frac{J}{M}}+\left({\frac{\alpha}{M}}\right)^2}}
        ,~\omega_i=0
        ~~\mbox{for} ~ \alpha > \alpha_R
    \end{eqnarray} 
with $\alpha_R \equiv \sqrt{JM}$. In the microcanonical ensemble
one may relate the average kinetic energy with the temperature:
$T/2=\langle p^2\rangle /2M=\alpha^2 /6M$, from which one has $T_R
=\alpha^2_R /3M =J/3$ \cite{com1}. Accordingly, it is concluded in
this case that the rotating solution is neutrally stable for $T >
T_R$ and becomes unstable below $T_R$. Note here that $T_R$ is
lower than the equilibrium critical temperature $T_c=J/2$.

For the antiferromagnetic system ($J<0$), we replace $J=-|J|$ in Eq.
(\ref{rof}) to obtain, for $\omega_i =0$,
\begin{equation}
    \omega_r
    =\pm\sqrt{\displaystyle{{\frac{|J|}{M}}+\left({\frac{\alpha}{M}}\right)^2}},
\end{equation}
while there is no solution for $\omega_r =0$. We therefore
conclude that the antiferromagnetic system is neutrally stable for all
$\alpha$, i.e., at all temperatures.  In Sec. \ref{sec:solfp} we have shown that
the bi-cluster state is allowed by the rotating distribution.
The result obtained here that this non-stationary solution is neutrally stable at all
temperatures thus suggests an alternative explanation as to the origin of the
spontaneously formed bi-cluster state in numerical simulations,
which retains its form for quite a long time~\cite{afrot}.
This keeps parallel with the emergence of quasi-stationarity in the
ferromagnetic system, associated with the neutral stability~\cite{cc}.

\section{Conclusion}\label{sec:con}

In this paper, we have presented a detailed analysis of the system
of globally coupled rotors. Starting from a set of Langevin
equations and their corresponding FPE, which includes the
microcanonical ensemble approach as a limiting case, we have found
a class of solutions and studied their stability.  The standard
canonical distribution constitutes a simultaneous solution of the
canonical and the microcanonical ensembles, and thus describes the
same equilibrium behavior in both ensembles, leading to the
coherent phase (characterized by a nonzero mono-cluster order
parameter, {i.e.}, $\Delta^{(1)} \ne 0$) below the critical
temperature $T_c$ in the ferromagnetic system. The stability of
the coherent phase is governed by an infinite-order difference
equation, the behavior of which may be understood by considering
successively higher-order terms (i.e., Fourier components in the
perturbation). It has been found that the coherent phase is stable
above some temperature $T_0$, which is finite if one includes only
a few lowest Fourier components. As more components are
considered, $T_0$ tends to decrease toward zero; this leads us to
surmise that the infinite number of Fourier components would
stabilize the coherent phase down to zero temperature. Namely, it
is expected that the stability equation, if treated exactly, leads
to the stability of the coherent phase at all temperatures below
$T_c$.

We find the more interesting possibility for the non-stationary
(rotating) solution with regard to dynamical order. Dynamical
order, manifested by multi-cluster motion, is allowed for both
ferromagnetic and antiferromagnetic interactions. Unlike a
ferromagnetic system, in which dynamical order ceases to exist
below the temperature $T_R$, dynamical order is observed to be
neutrally stable down to zero temperature in the antiferromagnetic
system. This suggests an alternative explanation as to the origin
of the spontaneous formation of the bi-cluster phase in the system
of antiferromagnetically coupled rotors. This is in parallel with
the explanation that the quasi-stationarity observed in
ferromagnetically coupled rotors is related to the neutral
stability of the stationary solution in the incoherent phase below
the equilibrium critical temperature~\cite{cc}.

To conclude, we have introduced a unified approach for both the
canonical ensemble and the microcanonical ensemble, based on the
Fokker-Planck equation. Depending on the ensemble, the
Fokker-Planck equation admits a few solutions which have
implications on some remarkable features (quasi-stationarity in
ferromagnetic systems and bi-cluster motion in antiferromagnetic
systems) observed in numerical experiments. We provide natural
explanations for the origin of those seemingly unrelated features
within the same context.
Finally, we point out that our approach is based on an effectively
one-particle dynamics, exact for an infinite number of particles
and does not reflect the instabilities that may be caused by the
finiteness of the number of particles.

\ack

JC thanks the Korea Institute for Advanced Study for hospitality during his stay,
where this work was completed.
This work was supported in part by the Korea Science and Engineering Foundation
through National Core Research Center for Systems Bio-Dynamics
and by the Ministry of Education through the BK21 Program.

\appendix*
\section{Properties of $\chi_k (\omega)$}

In this appendix we describe some properties of the response
function $\chi_k (\omega)$ for the Maxwell distribution $f_M (p)$:
\begin{equation}\label{ck}
    \chi_k (\omega)=\frac{J}{2}\int dp \frac{f'_M (p)}{\omega
    +kp/M}.
\end{equation}
First, for $k=0$, we have
\begin{equation}\label{c0}
    \chi_0 (\omega)=\frac{J}{2}\int dp \frac{f'_M (p)}{\omega}=0
\end{equation}
since $f'_M (p)$ is an odd function.
We next write $k\rightarrow -k $ and change the integration variable $p$ to $-p$
in Eq. (\ref{ck}) to get
\begin{eqnarray}
    \chi_{-k} (\omega)&=&\frac{J}{2}\int dp \frac{f'_M (p)}{\omega
    -kp/M} \nonumber \\
    &=&\frac{J}{2}\int d(-p) \frac{f'_M (-p)}{\omega
    +kp/M} \nonumber \\
    &=&-\frac{J}{2}\int dp \frac{f'_M (p)}{\omega
    +kp/M}= -\chi_{k} (\omega),
\end{eqnarray}
again noting that $f'_M (p)$ is an odd function. Similarly, it is
straightforward to show that
\begin{equation}\label{sign}
    \chi_k (-\omega)=-\chi_{-k}(\omega)=\chi_k (\omega).
\end{equation}
Further, Eq. (\ref{ck}) can also be written as
\begin{eqnarray}
    \chi_k (\omega)&=&\frac{JM}{2k}\int dp \frac{f'_M (p)}{p
    +M\omega /k}\nonumber \\
    &\equiv&\frac{1}{k} \chi(\omega /k),
\end{eqnarray}
where
\begin{equation}
\chi (\omega)\equiv \frac{JM}{2}\int dp \frac{f'_M (p)}{p
    +M\omega}=\chi(-\omega)
\end{equation}
is the response function already defined in Sec. {\ref{sec:sta}}.
Although we consider here the Maxwell distribution, the properties given above
hold for any momentum distribution $f_0 (p)$, only if it is an even function of $p$.
We now proceed to evaluate this function, paying attention to the
simple pole at $p=-M\omega$ on the complex $p$-plane.
Setting $M\omega \equiv \tilde{\omega}$ and making analytic continuation
$\omega= \omega_r +i \omega_i$, we obtain $\chi (\omega)$ in the form
\begin{eqnarray}
\tilde{\chi}(\omega)
  \equiv \frac{2}{JM}\chi(\omega) 
  = \left\{\begin{array}{ll}
     \int_{-\infty}^{\infty} dp \frac{\displaystyle{f'_M (p)}}
    {\displaystyle{p+ \tilde{\omega }}} &~\mbox{for} ~ \omega_i > 0 \\
     \mathcal{P}\int_{-\infty}^{\infty} dp \frac{\displaystyle{f'_M (p)}}
    {\displaystyle{p+ \tilde{\omega }}} -i\pi f'_M (-\tilde{\omega})&~\mbox{for} ~ \omega_i = 0 \\
     \int_{-\infty}^{\infty} dp \frac{\displaystyle{f'_M (p)}}
    {\displaystyle{p+ \tilde{\omega}}}-2i\pi f'_M (-\tilde{\omega}) &~\mbox{for} ~ \omega_i <
    0.
    \end{array}  \right.
\end{eqnarray}
With the tilde sign omitted for convenience, the real part reads
\begin{eqnarray}\label{rechi}
\mathrm{{Re}} \tilde{\chi}(\omega)
  &=& \left\{\begin{array}{ll}
     \int_{-\infty}^{\infty} dp \frac{\displaystyle{(p+\omega_r)f'_M
     (p)}}
    {\displaystyle{(p+ \omega_r )^2}+\omega^2_i}
     &~\mbox{for} ~ \omega_i > 0  \\
     P\int_{-\infty}^{\infty} dp \frac{\displaystyle{f'_M (p)}}
    {\displaystyle{p+ \omega_r }} +\pi \mathrm{Im}f'_M (-\omega)
     &~\mbox{for} ~ \omega_i = 0  \\
     \int_{-\infty}^{\infty} dp \frac{\displaystyle{(p+\omega_r)f'_M (p)}}
    {\displaystyle{(p+ \omega_r)^2 +\omega^2_i}}+2\pi \mathrm{Im}f'_M (-\omega)
     &~\mbox{for} ~ \omega_i < 0
    \end{array}  \right. 
\end{eqnarray}
while the imaginary part is given by
\begin{eqnarray}\label{imchi}
\mathrm{Im} \tilde{\chi}(\omega)
  &=& \left\{\begin{array}{ll}
     \omega_i \int_{-\infty}^{\infty} dp
     \frac{\displaystyle{f'_M(p)}}
    {\displaystyle{(p+ \omega_r )^2}+\omega^2_i} &~\mbox{for} ~ \omega_i > 0 \\
     \textrm{Re}f'_M (-\omega)&~\mbox{for} ~ \omega_i = 0 \\
     \omega_i \int_{-\infty}^{\infty} dp \frac{\displaystyle{f'_M (p)}}
    {\displaystyle{(p+ \omega_r)^2 +\omega^2_i}}+2\pi \mathrm{Re}f'_M (-\omega) &~\mbox{for} ~ \omega_i <
    0.
    \end{array}  \right. 
\end{eqnarray}
We next write
\begin{eqnarray}
  f'_M (\omega_r +i \omega_i) &=& \frac{\omega_r +i \omega_i}{\sqrt{2\pi M^3 T^3}}
  e^{-(\omega^2_r-\omega^2_i)/2MT} e^{-i \omega_r \omega_i/MT} \nonumber \\
   &=& -\frac{e^{-(\omega^2_r-\omega^2_i)/2MT}}{\sqrt{2\pi M^3 T^3}}
   \left[\omega_r \cos \frac{\omega_r \omega_i}{MT}+\omega_i \sin \frac{\omega_r \omega_i}{MT} \right. \nonumber \\
   & &~~~~~~~~~~~~~~~\left.+i\left(\omega_i \cos \frac{\omega_r \omega_i}{MT}-\omega_r \sin \frac{\omega_r
   \omega_i}{MT}\right)\right] \nonumber \\
   &\equiv & \mathrm{Re} f'_M (\omega_r{+}i\omega_i)
           + i\mathrm{Im} f'_M (\omega_r{+}i\omega_i),
\end{eqnarray}
from which it is obvious that $\mathrm{Re} f'_M (\omega_r +i\omega_i)=0$ for $\omega_r =0$ and
$\mathrm{Im} f'_M (i\omega_i)=-(2\pi M^3 T^3)^{-1/2} \omega_i e^{\omega^2_i/2MT}$.
We thus conclude that $\omega_r =0$ is a solution of $\mathrm{Im}
\tilde{\chi}(\omega)=0$, since $f'_M (p)$ is an odd function of $p$,
which makes the integrals vanish in Eq. (\ref{imchi}).
We now evaluate the integral of $\mathrm{Re} \tilde{\chi}(\omega)$.
For $\omega_r >0$, the first equation in Eq. (\ref{rechi}) becomes \cite{as}
\begin{eqnarray}\label{iplus}
    \textrm{Re} \tilde{\chi}(\omega)&=&-\frac{1}{T}[1-\sqrt{\pi}ye^{y^{2}}\textrm{erfc}
    (y)] \nonumber \\
    &\equiv&-\frac{1}{T}f(y)
\end{eqnarray}
with the scaled variable $y\equiv \omega_i \sqrt{M/2T}$, where
\begin{equation}\label{iminus}
    \textrm{erfc}(y)=\frac{2}{\sqrt{\pi}}\int_y^{\infty}e^{-t^2}dt
\end{equation}
is the complimentary error function.
For $\omega_i =-|\omega_i| <0$, it is straightforward to show that the last equation
in Eq. (\ref{rechi}) becomes
\begin{eqnarray}\label{izero}
    \mathrm{Re} \tilde{\chi}(\omega)&=&-\frac{1}{T}[1+\sqrt{\pi}|y|e^{y^{2}}(2-\mathrm{erfc}
    (|y|))] \nonumber \\
    &\equiv&-\frac{1}{T}g(y).
\end{eqnarray}
For $\omega_i =0$, we have $\mathrm{Im} f'_M (\omega_r +i \omega_i)=0$ and
the second equation in Eq. (\ref{rechi}) simply reduces to
$\mathrm{Re} \tilde{\chi}(\omega) =-1/T$.
Note that $f(y)$ is a monotonically decreasing function of $y$, varying from unity to zero
as $y$ grows from zero to arbitrarily large values.
On the other hand, $g(y)$ increases monotonically with $y$, from unity to arbitrarily large values.


\begin{thebibliography}{99}
\bibitem{period}
Here the summation of the infinite series in Eq. (\ref{non-sta}) yields
$P^{(0)}(\phi,p,t)= 2\pi F(p)\delta (2(\phi{-}pt/M))$ with the
$2\pi$-periodicity in the argument.  From this equation, manifesting the
periodicity, we obtain Eq. (\ref{pi1}).
Similarly, Eq. (\ref{pi2}) is obtained.
\bibitem{comph}
In fact, there is phase difference of $\theta_\ell -\theta_1$ for
arbitrary $\ell$, which may be disregarded. For the incoherent
phase the global phase $\theta_1$ does not come in.
\bibitem{com1}
Numerical experiments usually use the internal energy $U$ as the
control parameter, which is related with the temperature $T$ via
$U=T/2 + (1-\Delta^2)/2$ for $J=1$. We here use a slightly
different definition of the potential energy.
\end{thebibliography}
\end{document}